\documentclass{ws-p8-50x6-00}
\begin{document}

\title{Interactions, Currents, and the Structure of Few-Nucleon
Systems}

\author{R.\ Schiavilla}

\address{Jefferson Lab, Newport News, VA 23606 \\
and \\
Old Dominion University, Norfolk, VA 23529}

\maketitle

\abstracts{Our current understanding of the structure of
nuclei with $A \leq 8$, including energy spectra, electromagnetic
form factors, and weak transitions, is reviewed within the context of a
realistic approach to nuclear dynamics based on two- and three-
nucleon interactions and associated electro-weak currents.
Low-energy radiative and weak capture reactions of
astrophysical relevance involving these light systems are also discussed.}

\section{Introduction}

Few-nucleon systems provide a unique opportunity for testing
the simple, traditional picture of the nucleus as a system of
point-like nucleons interacting among themselves via effective
many-body potentials, and with external electro-weak probes
via effective many-body currents.  Through advances in computational
techniques and facilities, the last few years have witnessed dramatic
progress in numerically exact studies of the structure and
dynamics of systems with mass number $A \leq 8$, including
energy spectra of low-lying states, momentum distributions
and cluster amplitudes, elastic and inelastic electromagnetic
form factors, $\beta$-decays, radiative and weak capture
reactions at low energies, inclusive response to hadronic and
electro-weak probes at intermediate energies.  

In the present talk, I will review the \lq\lq nuclear standard
model\rq\rq outlined above, and present the extent to which it is
successful in predicting some of the nuclear properties alluded
to earlier.  Of course, given the limited time, some of the theoretical
and experimental developments will be treated cursorily.
Nevertheless, I still hope to be able to convey a broad view
of the intriguing and important studies in few-nucleon physics
today.
\section{Potentials and Energy Spectra}

The Hamiltonian in the nuclear standard model is written as

\begin{equation}
H= \sum_i K_i + \sum_{i<j} v_{ij} +\sum_{i<j<k} V_{ijk} \ ,
\end{equation}
where the kinetic energy operator $K_i$ has charge-independent
and charge-symmetry-breaking components due to the difference
in proton and neutron masses, and $v_{ij}$ and $V_{ijk}$ are two- and
three-nucleon potentials.  

The two-nucleon potential consists of a long range part due to
pion exchange, and a short-range part parameterized either
in terms of heavy meson exchanges as, for example, in
the Bonn potential~\cite{Bonn},
or via suitable operators and strength functions, as in the Argonne $v_{18}$
(AV18) potential~\cite{WSS95}.  The short-range terms in these potentials are then
constrained to fit $pp$ and $np$ scattering data up to energies of
$\simeq$ 350 MeV in the laboratory, and the deuteron binding energy.  The modern
models mentioned above provide fits to the Nijmegen data-base~\cite{Nijm} characterized
by $\chi^2$ per datum very close to one, and should therefore be viewed
as phase-equivalent.  The AV18 model is most widely
used; it has the form 

\begin{eqnarray}
v_{ij}&=&v_{ij}^\pi+v_{ij}^R \nonumber \\
      &=& \sum_{p=1,18} v^p(r_{ij}) \, O^p_{ij} \ ,
\end{eqnarray}
where the first fourteen operators are isoscalar,

\begin{equation}
O^{p=1-14}_{ij} = \left [ 1 , \sigma_i\cdot \sigma_j  ,
   S_{ij} ,   ({\bf L}\cdot{\bf S})_{ij} ,  {\bf L}^2  ,
  {\bf L}^2\sigma_i\cdot \sigma_j  ,
({\bf L} \cdot {\bf S})_{ij}^2  \right ]
\otimes \left [  1 ,  \tau_i\cdot \tau_j \right ] \ ,
\end{equation}
while the last four isospin-symmetry-breaking operators have isovector
and isotensor character,

\begin{equation}
O^{p=15-18}_{ij} = T_{ij}\, ,\,  \sigma_i\cdot \sigma_j T_{ij}\, ,
\,  S_{ij} T_{ij}\, , (\tau_i+\tau_j)_z \ .
\end{equation}
Here $S_{ij}$ is the tensor operator, and $T_{ij}$ is defined as
$T_{ij}=3\tau_{iz} \tau_{jz} -\tau_i \cdot \tau_j$.  Unique among the
modern potentials, the AV18 includes a fairly complete
treatment of the electromagnetic interaction, since it retains, in addition
to the leading Coulomb term, also contributions from magnetic moment
interactions, vacuum polarization and two-photon exchange corrections.
These terms, while typically very small (for example, in the deuteron
the magnetic dipole-dipole interaction gives 18 keV extra repulsion~\cite{WSS95}),
need to be taken into account when very accurate predictions are required, 
as in the case, for example, of studies of energy differences of isomultiplet
states~\cite{Wir00}, or the cross section for proton weak capture on
proton at keV energies~\cite{Sch98}.
 
It is now well established that two-nucleon potentials alone underbind
nuclei~\cite{Wir00}: for example, the AV18 and Bonn models give~\cite{NKG00}, in
numerically exact calculations, binding energies of 24.28 MeV and 26.26 MeV
respectively, which should be compared to the experimental value
of 28.3 MeV.  Moreover, $^6$Li and $^7$Li are unstable against breakup into $\alpha$$d$ and
$\alpha$$t$ clusters, respectively, and that energy differences are not, in
general, well predicted, when only two-nucleon potentials are retained
in the Hamiltonian.

Important components of the three-nucleon potential arise from the internal structure
of the nucleon.  Since all degrees of freedom other than the nucleon 
have been integrated out, the presence of virtual $\Delta$ resonances, for
example, induces three-nucleon potentials.  They are written as

\begin{equation}
V_{ijk}=V^{2\pi}_{ijk}+V^R_{ijk} \ ,
\end{equation}
where $V^{2\pi}$ is the \lq\lq long-range\rq\rq term, resulting from
the intermediate excitation of a $\Delta$ with pion exchanges involving the
other two nucleons, known as the Fujita-Miyazawa term~\cite{FM57}.  This term
is present in all models, such as the Tucson-Melbourne potential~\cite{Coon} or the
series of Urbana models~\cite{Pud95}.  The Urbana models parameterize $V^R$ as

\begin{equation}
V^R=A^R \sum_{{\rm cyclic}\> ijk} T^2_\pi(r_{ij}) T^2_\pi(r_{jk}) \ ,
\end{equation}
where $T_\pi(r)$ is the strength function of the pion-exchange tensor
interaction.  This term is meant to simulate the dispersive effects
that are required when integrating out $\Delta$ degrees of freedom.
The strengths of the Fujita-Miyazawa and dispersive terms are then
determined, in the Urbana models, by fitting the triton binding energy
and the saturation density of nuclear matter.  

The Hamiltonian consisting of the AV18 two-nucleon and Urbana-IX
three-nucleon potentials (AV18/UIX) predicts reasonably well the low-lying energy
spectra of systems with $A \leq 8$ nucleons in \lq\lq exact\rq\rq
Green's function Monte Carlo calculations~\cite{Wir00}.  The experimental
binding energies of the $\alpha$ particle is exactly
reproduced, while those of the $A$=6--8 systems are underpredicted
by a few percent.  This underbinding becomes (relatively) more and more
severe as the neutron-proton asymmetry increases.  An additional failure
of this Hamiltonian model is the underprediction of spin-orbit splittings
in the excitation spectra of these light systems.  These failures
have in fact led to the development of new three-nucleon interaction
models~\cite{Car00}.  These newly developed models, denoted as Illinois
models, incorporate the Fujita-Miyazawa and dispersive terms
discussed above, but include in addition multipion exchange terms
involving excitation of one or two $\Delta$'s, so-called pion-ring
diagrams, as well as the terms arising from S-wave pion rescattering,
required by chiral symmetry. 
\section{The Nuclear Electromagnetic Current}

The nuclear current operator consists of one- and many-body terms
that operate on the nucleon degrees of freedom:

\begin{equation}
{\bf j}({\bf q})= \sum_i {\bf j}^{(1)}_i({\bf q})
             +\sum_{i<j} {\bf j}^{(2)}_{ij}({\bf q})
             +\sum_{i<j<k} {\bf j}^{(3)}_{ijk}({\bf q}) \ ,
\end{equation}
where ${\bf q}$ is the momentum transfer, and
the one-body operator ${\bf j}^{(1)}_i$ has the
standard expression in terms of single-nucleon
convection and magnetization currents.  The two-body current
operator has \lq\lq model-independent\rq\rq and
\lq\lq model-dependent\rq\rq components (for a review, see Ref.~\cite{Car98}).
The model-independent terms are obtained from the charge-independent part
of the AV18, and by construction satisfy current conservation
with this interaction.  The leading operator is the isovector
\lq\lq $\pi$-like\rq\rq current obtained
from the isospin-dependent spin-spin and tensor interactions.
The latter also generate an isovector \lq\lq $\rho$-like \rq\rq current, while
additional model-independent isoscalar and isovector currents arise from the
central and momentum-dependent interactions.  These currents are short-ranged
and numerically far less important than the $\pi$-like current.  Finally,
models for three-body currents have been derived in Ref.~\cite{Mar98}, however
the associated contributions have been found to be very small in studies
of the magnetic structure of the trinucleons~\cite{Mar98}.

The model-dependent currents are purely transverse
and therefore cannot be directly linked to the underlying
two-nucleon interaction.  Among them, those associated with
the $\Delta$-isobar are the most important ones in the
momentum-transfer regime being discussed here.  These currents are
treated within the transition-correlation-operator
(TCO) scheme~\cite{Mar98,Sch92}, a scaled-down
approach to a full $N$+$\Delta$ coupled-channel treatment.
In the TCO scheme, the $\Delta$ degrees of freedom
are explicitly included in the nuclear wave functions by writing

\begin{equation}
\Psi_{N+\Delta}=\left[{\cal{S}}\prod_{i<j}\left(1\,+\,U^{TR}_{ij}\right)
\right]\, \Psi \ ,
\label{eq:psiNDtco}
\end{equation}
where $\Psi$ is the purely nucleonic component, $\cal{S}$ is the
symmetrizer and the transition correlations $U^{TR}_{ij}$ are
short-range operators, that convert $NN$ pairs
into $N\Delta$ and $\Delta\Delta$ pairs.
In the results reported here, the $\Psi$
is taken from CHH solutions of the AV18/UIX Hamiltonian with nucleons only
interactions, while the $U^{TR}_{ij}$ is obtained from two-body
bound and low-energy scattering state solutions of the full $N$-$\Delta$
coupled-channel problem.  Both $\gamma N \Delta$ and $\gamma \Delta \Delta$
$M_1$ couplings are considered with their values,
$\mu_{\gamma N \Delta}=3$ n.m.\ and
$\mu_{\gamma \Delta \Delta}=4.35$ n.m., obtained from data~\cite{Sch92}.
\section{The $p$$d$ Radiative Capture}

There are now available many high-quality data, including differential
cross sections, vector and tensor analyzing powers, and photon polarization
coefficients, on the $pd$ radiative capture at c.m.\ energies
ranging from 0 to 2 MeV~\cite{Sea96,Mea97,Wea99,SK99}.  These data indicate
that the reaction proceeds predominantly through S- and P-wave capture.
The aim here is to verify the extent to which they can be described
satisfactorily by a calculation based on a realistic Hamiltonian
(the AV18/UIX model) and a current operator constructed consistently
with the two- and three-nucleon interactions~\cite{Viv00}.
\begin{figure}[bth]
\centerline{
\epsfig{file=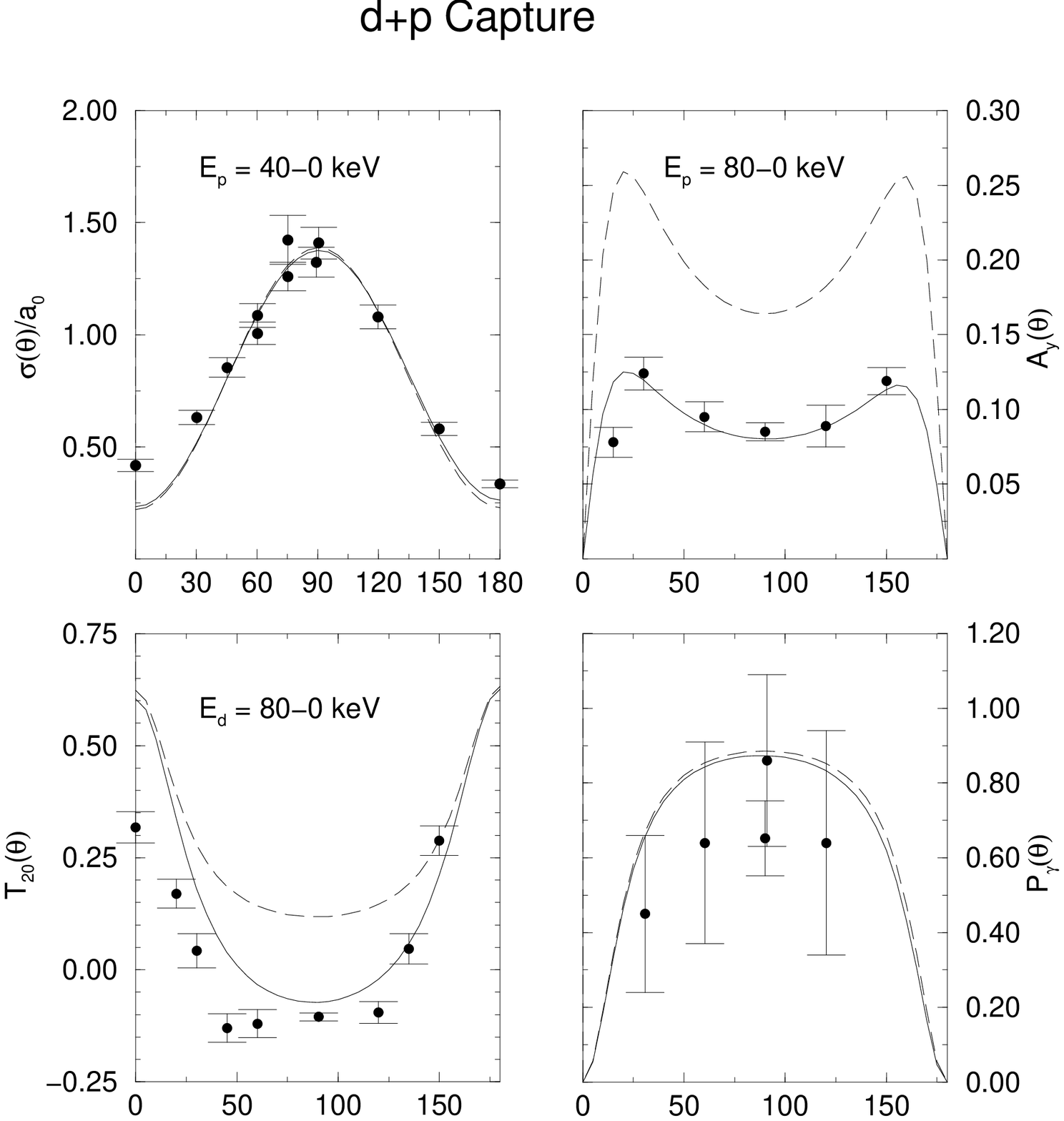,height=2.5in,width=3.5in}}
\caption{The energy integrated cross section $\sigma(\theta)/a_0$
($4\pi a_0$ is the total cross section), vector analyzing power
$A_y(\theta)$, tensor analyzing power $T_{20}(\theta)$ and photon
linear polarization coefficient $P_\gamma(\theta)$ obtained with the
AV18/UIX Hamiltonian model and one-body only (dashed line)
or both one- and many-body (solid line) currents
are compared with the experimental results of Ref.~\protect\cite{Sea96}.}
\label{fig:cpt080keV}
\end{figure}

The predicted angular distributions of the differential
cross section $\sigma(\theta)$, vector and tensor analyzing powers
$A_y(\theta)$ and $T_{20}(\theta)$, and photon
linear polarization coefficient $P_\gamma(\theta)$ are compared with
the TUNL data below 50 keV from
Refs.~\cite{Sea96,Wea99} in Fig~\ref{fig:cpt080keV}.
The agreement between the full theory, including many-body
current contributions, and experiment is generally good.
However, a closer inspection of the figure reveals the
presence of significative discrepancies between theory and experiment
in the small angle behavior of $\sigma(\theta)$ and $T_{20}(\theta)$, as well
as in the $S$-factor below 40 keV~\cite{Viv00}.
The S-wave capture proceeds mostly through the $M_1$ transitions connecting
the doublet and quartet $pd$ states to $^3$He--the associated reduced
matrix elements (RMEs) are denoted by $m_2$ and $m_4$, respectively.
The situation for P-wave capture is more complex, although at energies
below 50 keV it is dominated by the $E_1$ transitions
from the doublet and quartet $pd$ states having channel spin
$S$=1/2, whose RMEs I denote as $p_2$ and $p_4$.  The $E_1$
transitions involving the channel spin $S=3/2$ states, while
smaller, do play an important role in $T_{20}(\theta)$.

The TUNL~\cite{Wea99} and Wisconsin~\cite{SK99} groups have determined
the leading $M_1$ and $E_1$ RMEs via fits to the measured observables.
The results of this fitting procedure are compared with the
calculated RMEs in Table~\ref{tab:rmew}.
The phase of each RME is simply related to
the elastic $pd$ phase shift~\cite{SK99}, which at these low energies
is essentially the Coulomb phase shift.
As can be seen from Table~\ref{tab:rmew},
the most significant differences between theoretical and experimental RMEs
are found for $|p_4|$.  The theoretical overprediction of $p_4$
is the cause of the discrepancies mentioned above in the
low-energy ($\le 50$ keV) $S$-factor and small
angle $\sigma(\theta)$.

It is interesting to analyze the ratio
$r_{E1} \equiv |p_4/p_2|^2$.
Theory gives $r_{E1}\simeq 1$, while from
the fit it results that $r_{E1}\approx 0.74\pm0.04$.
It is important to stress that the calculation of these RMEs is not
influenced by uncertainties in the two-body currents, since their
values are entirely given by the long-wavelength form of the $E_1$
operator (Siegert's theorem), which has no spin-dependence (for a thorough
discussion of the validity of the long-wavelength approximation
in $E_1$ transitions, particularly suppressed
ones, see Ref.~\cite{Viv00}).  It is therefore
of interest to examine more closely the origin of the above discrepancy.
If the interactions between the $p$ and $d$ clusters are switched off,
the relation $r_{E1} \simeq 1$ then simply follows
from angular momentum algebra.
Deviations of this ratio from one are therefore
to be ascribed to differences induced by the interactions in the
$S$=1/2 doublet and quartet wave functions.
The AV18/UIX interactions in these channels do not change
the ratio above significantly.  It should be emphasized that
the studies carried out up until now ignore, in the continuum states, the
effects arising from electromagnetic interactions
beyond the static Coulomb interaction between protons.
It is not clear whether the inclusion of these
long-range interactions, in particular their spin-orbit component,
could explain the splitting between
the $p_2$ and $p_4$ RMEs observed at very low energy.
This discrepancy seems to disappear at 2 MeV~\cite{Viv00}.
\begin{table}[htb]
\caption{Magnitudes of the leading $M_1$ and $E_1$ RMEs
for $pd$ capture at $E_p=40$ keV.}
\label{tab:rmew}
\newcommand{\m}{\hphantom{$-$}}
\newcommand{\cc}[1]{\multicolumn{1}{c}{#1}}
\renewcommand{\tabcolsep}{1.5pc} 
\begin{tabular}{@{}llll}
RME & IA & FULL & FIT \\
\hline
 $|m_2|$ & 0.172  & 0.322 & 0.340$\pm$0.010  \\
 $|m_4|$ & 0.174  & 0.157 & 0.157$\pm$0.007  \\
 $|p_2|$ & 0.346  & 0.371 & 0.363$\pm$0.014  \\
 $|p_4|$ & 0.343  & 0.378 & 0.312$\pm$0.009  \\
\hline
\end{tabular}\\[2pt]
\end{table}

Finally, the doublet $m_2$ RME is underpredicted by theory at the
5 \% level.  On the other hand, the cross section for $n$$d$
capture at thermal neutron energy is calculated to be 578 $\mu$b
with the AV18/UIX model, which is 15 \% larger than the experimental
value (508$\pm$15) $\mu$b~\cite{JBB82}.
Of course, $M_1$ transitions, particularly
doublet ones, are significantly influenced by many-body current
contributions.  Indeed, an analysis of the
isoscalar ($\mu_S$) and isovector ($\mu_V$)
magnetic moments of the trinucleons~\cite{Mar98}
suggests that the present model
for the isoscalar two-body currents, constructed from the AV18 spin-orbit
and quadratic-momentum dependent interactions, tends to overestimate
$\mu_S$ by about 5 \%.  The experimental value for $\mu_V$, however, is almost
perfectly reproduced.  It appears that the present model for two-body
currents needs to be improved.
\section{The Nuclear Weak Current and the $p\,^3$He Weak Capture}

The nuclear weak current and charge operators
consist of vector and axial-vector parts, with
corresponding one- and many-body components.  The weak
vector current and charge
are constructed from the corresponding (isovector)
electromagnetic terms, in accordance with the
conserved-vector-current hypothesis, and thus
have~\cite{Mar00} \lq\lq model-independent\rq\rq
and \lq\lq model-dependent\rq\rq components.  The former
are determined by the interactions, the latter
include the transverse currents associated with $\Delta$ excitation.

The leading many-body terms in the
axial current, in contrast to the case of the
weak vector (or electromagnetic) current, are those due to $\Delta$
excitation, which are treated within the TCO scheme, discussed above.
The axial charge operator includes
the long-range pion-exchange term~\cite{Kub78}, required by low-energy
theorems and the partially-conserved-axial-current relation,
as well as the (expected) leading short-range terms constructed
from the central and spin-orbit components of the nucleon-nucleon
interaction~\cite{Kir92}.

The largest model dependence is in the weak axial current.
The $N$$\Delta$ axial coupling constant
$g_{A}^{*}$ is not well known.  In the quark-model, it is
related to the axial coupling constant
of the nucleon by the relations $g_{A}^{*}=(6\sqrt{2}/5) g_A$.
This value has often been used in the
literature in the calculation of $\Delta$-induced axial current contributions
to weak transitions.  However, given the uncertainties inherent
to quark-model predictions, a more reliable estimate for $g_A^*$
is obtained by determining its value phenomenologically.
It is well established by now~\cite{Sch98}
that one-body axial currents
lead to a $\simeq$ 4 \% underprediction of the measured
Gamow-Teller matrix element in tritium $\beta$-decay.
This small 4 \% discrepancy can then be used
to determine $g_{A}^{*}$~\cite{Mar00}.  While this
procedure is inherently model dependent, its actual
model dependence is in fact very weak, as has been
shown in Ref.~\cite{Sch98}.

The calculated values for the astrophysical $S$-factor
in the energy range 0--10 keV
are listed in Table~\ref{tb:sfact}~\cite{Mar00}.  Inspection of the
table shows that: (i) the energy dependence is
rather weak, the value at $10$ keV is only about 4 \% larger
than that at $0$ keV; (ii) the P-wave capture states are found to
be important, contributing about 40 \% of the calculated
$S$-factor.  However, the contributions from D-wave channels
are expected to be very small, as explicitly
verified in $^3$D$_1$ capture.
(iii) The many-body axial currents
associated with $\Delta$ excitation play a crucial
role in the (dominant) $^3$S$_1$ capture, where they reduce
the $S$-factor by more than a factor of four; thus
the destructive interference between the one- and many-body
current contributions, obtained in Ref.~\cite{Sch92}, is
confirmed in the study of Ref.~\cite{Mar00}, based on more accurate
wave functions.  The (suppressed) one-body contribution
comes mostly from transitions involving the D-state
components of the $^3$He and $^4$He wave functions, while
the many-body contributions are predominantly due to transitions
connecting the S-state in $^3$He to the D-state in $^4$He, or viceversa.
\begin{table}[htb]
\caption{The $hep$ $S$-factor, in units of $10^{-20}$ keV~b, calculated
with CHH wave functions corresponding to the AV18/UIX Hamiltonian model,
at $p\,^3$He c.m.\ energies $E$=0, 5, and 10 keV.  The rows
labelled \lq\lq one-body\rq\rq and \lq\lq full\rq\rq list the
contributions obtained by retaining the one-body only and both
one- and many-body terms in the nuclear weak current.  The contributions due
the $^3$S$_1$ channel only and all S- and P-wave channels are
listed separately.}
\label{tb:sfact}
\newcommand{\m}{\hphantom{$-$}}
\newcommand{\cc}[1]{\multicolumn{1}{c}{#1}}
\renewcommand{\tabcolsep}{1.2pc} 
\begin{tabular}{@{}lllllll}
\hline
& \multicolumn{2}{c} {$E$=$0$ keV} &
  \multicolumn{2}{c} {$E$=$5$ keV} &
  \multicolumn{2}{c} {$E$=$10$ keV} \\
\hline
& $^3$S$_1$ & S+P & $^3$S$_1$ & S+P & $^3$S$_1$ & S+P\\
\hline
one-body  &26.4  & 29.0 & 25.9 & 28.7 & 26.2 & 29.3 \\
full      &6.38  & 9.64 & 6.20 & 9.70 & 6.36 & 10.1 \\
\hline
\end{tabular}
\end{table}

The chief conclusion of Ref.~\cite{Mar00} is
that the $hep$ $S$-factor is predicted to be $\simeq$ 4.5
times larger than the value adopted in the
standard solar model (SSM)~\cite{BBP98}.  This enhancement,
while very significant, is smaller than that first suggested
in Ref.~\cite{BK98}.  Even though this result is
inherently model dependent, it is unlikely that the model dependence
is large enough to accommodate a drastic increase in the value obtained here.
Indeed, calculations using Hamiltonians based on the AV18 two-nucleon
interaction only and the older AV14/UVIII two- and
three-nucleon interactions~\cite{WSA84} predict zero energy $S$-factor values of
$12.1 \times 10^{-20}$ keV~b and $10.2 \times 10^{-20}$ keV~b, respectively.
It should be stressed, however, that the AV18 model, in contrast to the
AV14/UVIII, does not reproduce the experimental binding energies and low-energy
scattering parameters of the three- and four-nucleon systems.
The AV14/UVIII prediction is only 6 \% larger than the AV18/UIX
zero-energy result.  This 6 \% variation should provide
a fairly realistic estimate of the theoretical uncertainty due to
the model dependence.  The precise calculation of the $S$-factor and the consequent
absolute prediction for the $hep$ neutrino flux should allow much
greater discrimination among proposed solar neutrino oscillation
solutions~\cite{Mar00}.
\section{Conclusions and Acknowledgments}

Improvements in the modeling of two- and three-nucleon interactions and
nuclear electro-weak currents, and the significant
progress made in the last few years in the description of bound and
continuum wave functions, make it now possible
to perform first-principle calculations of interesting nuclear
properties of light nuclei.  Experimentally known electromagnetic and weak transitions
of systems in the mass range $2 \leq A \leq 8$ provide
powerful constraints on models of nuclear currents.

I wish to thank J.\ Carlson, A.\ Kievsky, L.E.\ Marcucci, V.R.\ Pandharipande,
S.C.\ Pieper, D.O.\ Riska, S.\ Rosati, M.\ Viviani, and R.B.\ Wiringa
for their many important contributions to the work reported here.
This work was supported by DOE contract DE-AC05-84ER40150 under which the
Southeastern Universities Research Association (SURA) operates
the Thomas Jefferson National Accelerator Facility.
\end{document}